\documentclass[prd,preprint,nofootinbib]{revtex4}
\usepackage{amssymb}
\usepackage{epsfig}

\begin{document}

\title{Invisible Technicolor}

\author{Naoyuki Haba}
\email{haba@ias.tokushima-u.ac.jp}
\affiliation{
Institute of Theoretical Physics, University of Tokushima,
Tokushima 770-8502, Japan}

\author{Noriaki Kitazawa}
\email{kitazawa@phys.metro-u.ac.jp}
\affiliation{
Department of Physics, Tokyo Metropolitan University,
Hachioji, Tokyo 192-0397, Japan}

\author{Nobuchika Okada}
\email{okadan@post.kek.jp}
\affiliation{
Theory Division, KEK, Tsukuba, Ibaraki 305-0801, Japan}

\date{May 18, 2005}

\begin{abstract}
We propose a scenario of
 dynamical electroweak symmetry breaking
 which is applicable to the model with
 no tree-level potential for the elementary Higgs doublet field.
An example of such a model
 is the gauge-Higgs unification model.
The strong coupling ``technicolor'' dynamics
 can provide the scale of the electroweak symmetry breaking
 in the potential of the Higgs doublet field.
The negative mass squared and quartic coupling can be generated
 through the Yukawa couplings among heavy and light
 ``technifermions'' and the Higgs doublet field.
Since the massless ``technifermion'' is singlet
 under the electroweak gauge symmetry,
 no large corrections to the electroweak observables arise. 
As a prediction of this scenario,
 there must be a pseudo-Nambu-Goldstone boson
 which couple with the Higgs field in a specific way,
 though it is singlet under the standard model gauge symmetry.
\end{abstract}

\pacs{}
\preprint{KEK-TH-1015}
\vspace*{2cm}
\maketitle

\section{Introduction}
\label{sec:introduction}

The dynamics of the electroweak symmetry breaking is still unknown.
In the standard model
 the electroweak symmetry is spontaneously broken
 by the vacuum expectation value of an elementary scalar field,
 the Higgs doublet field.
The energy scale of the symmetry breaking
 is set by hand as the mass of the Higgs doublet field.
This scenario may be true,
 but it is probable that some unknown dynamics
 determine the energy scale of the symmetry breaking.

Technicolor theory has been proposed as a scenario in which
 the energy scale is dynamically given
 by the pair condensation of the technifermion
 due to the strong coupling technicolor gauge interaction
 \cite{Weinberg,Susskind}.
The technicolor theory has a beautiful concept:
 all the matter are fermion fields
 and all the interactions are originated from gauge symmetry.
But, it is well known that
 some special dynamics or mechanisms are required
 so that this scenario is consistent with
 the precision electroweak measurements
 \cite{Peskin-Takeuchi,Sundram-Hsu,Kitazawa-Yanagida,
       Appelquist-Sannino,HHS}.
The problem is that
 the strong coupling technicolor dynamics
 directly affects the electroweak interaction.

Gauge-Higgs unification is a scenario
 which has the similar concept of the technicolor theory
 \cite{Manton,Fairlie,Hosotani,Hall-Nomura-Smith,Burdman-Nomura}.
The Higgs doublet field
 is originated from a gauge boson
 in the five-dimensional space-time,
 and no elementary scalar fields are required.
Because of the gauge symmetry,
 there is no Higgs potential at tree level,
 and the finite radiative correction may give
 non-trivial potential and trigger electroweak symmetry breaking
 \cite{Kudo-Lim-Yamashita,HHKY}.

In this letter
 we propose a scenario of the electroweak symmetry breaking
 in the model in which
 the tree-level potential of the Higgs doublet field
 is suppressed by some mechanism.
Gauge-Higgs unification models are good example of such models
 \footnote{
  Little Higgs models\cite{little-Higgs}
  also can be good examples}.
We introduce a strong coupling gauge interaction, ``technicolor'',
 but the resultant fermion pair condensation
 does not directly break the electroweak symmetry,
 since the fermion, ``technifermion'',
 is singlet under the standard model gauge group.
The strong coupling gauge dynamics
 delivers electroweak symmetry breaking scale
 to the potential of the Higgs doublet field
 through the Yukawa couplings among
 Higgs doublet field and technifermions
 \footnote{
 The similar mechanism of the dynamical electroweak symmetry breaking
 through the vacuum misalignment by the Yukawa coupling
 has been proposed in Ref.\cite{Kaplan-Georgi}.
 The combination of
 the massive (effectively) elementary Higgs doublet field
 and the dynamical electroweak symmetry breaking
 by the usual technicolor has been considered in Ref.\cite{Choi-Kim}.
 }.
Since the scale is determined by the technicolor dynamics,
 it is natural to have small electroweak breaking scale
 in comparison with the Planck scale.
Some composite states, especially a pseudo-Nambu-Goldstone bosons,
 may couple to the Higgs field through the Yukawa couplings
 and may give some observable effects
 in future high-energy colliders.

In the following
 we discuss the scenario based on gauge-Higgs unification models
 for concreteness, though it is applicable to any models with
 no tree-level potential of the Higgs doublet field.

\section{The Dynamics}
\label{sec:dynamics}

In the gauge-Higgs unification scenario,
 some appropriate large gauge symmetry
 which includes standard model gauge symmetry
 is assumed in five-dimensional space-time.
The gauge symmetry
 is broken to the standard model gauge symmetry
 through the compactification of the fifth dimension.
The fifth component
 of some gauge boson of the original gauge symmetry
 can be a Higgs doublet field which is a scalar field
 in our non-compact four-dimensional space-time.
There is no tree-level scalar potential
 for such Higgs doublet field due to the original gauge symmetry.
Since the original gauge symmetry
 is broken by the compactification of the fifth dimension,
 quantum corrections give a non-trivial
 potential to the Higgs doublet field.
In fact,
 the one-loop contribution of gauge bosons to the potential
 has been calculated in some concrete models
 \cite{Kudo-Lim-Yamashita,HHKY},
 and typically the potential is bounded from below
 around zero vacuum expectation value of the Higgs doublet field.
It means that
 the gauge boson one-loop contribution
 to the Higgs mass squared is typically positive,
 and some physics is required to have electroweak symmetry breaking.
One of such physics is
 to introduce appropriate number of (massive) matter fields
 in the five-dimensional space-time (bulk),
 which gives additional contribution to the Higgs potential.
Typically,
 the contribution to the Higgs mass squared
 is negative for bulk fermions and positive for bulk bosons
 \footnote{
  It it interesting to note that
   the one-loop contribution of the top quark
   with large Yukawa coupling
   solely can give realistic value of the Higgs mass squared.
 }.
Therefore,
 by arranging the field contents in the bulk
 it is possible to realize the electroweak symmetry breaking.
Of course,
 we also have to take care of
 the higher power terms of the Higgs potential
 to have realistic value of the vacuum expectation value
 of the Higgs doublet field.

Now we point out that another contribution is possible
 in case that we have strong coupling gauge interactions.
Suppose SU$(N_{TC})$ gauge symmetry
 ($N_{TC} > 2$), ``technicolor'', in the bulk
 with appropriate fermion field contents
 which give the following field contents
 in four-dimensional effective theory
 \footnote{
 For example,
  in the models based on the bulk gauge symmetry
  SU$(6) \supset$ SU$(3)_c \times$SU$(2)_L \times$U$(1)_Y$
  \cite{Hall-Nomura-Smith,Burdman-Nomura},
  this field contents and the Yukawa coupling of
  Eq.(\ref{Yukawa-interaction}) can be obtained
  by introducing two bulk fermion fields
  which belong to the fundamental representation of SU$(6)$
  and also to the fundamental representation of SU$(N_{TC})$.
 The ${\bf Z}_2$ orbifold parity should be opposite
  in these two fermion fields.
 }.
\begin{center}
\begin{tabular}{cccc}
 & \ SU$(N_{TC})$ \ & \ SU$(2)_L$ \ & \ U$(1)_Y$ \ \\
$\chi_{L,R}$ & $N_{TC}$ & $2$ & $-1/2$ \\
$\psi_{L,R}$ & $N_{TC}$ & $1$ & $0$ \\
\label{technifermions}  
\end{tabular}
\end{center}
Suppose also that 
 the field $\chi$ has Dirac mass $M$
 which is smaller than the compactification scale
 $1/R \equiv \Lambda$.
It is possible that these ``technifermions''
 interact with Higgs doublet field $\Phi$
 through the following Yukawa couplings.
\begin{equation}
 {\cal L}_{TC}^{\rm Yukawa}
  = g_2 \left( {\bar \psi} \chi \Phi
             + {\bar \chi} \psi \Phi^\dag
        \right),
\label{Yukawa-interaction}
\end{equation}
 where $g_2$ is the SU$(2)_L$ gauge coupling constant.
The SU$(N_{TC})$ interaction
 is asymptotically free in the four-dimensional effective theory
 and the coupling becomes strong at the scale $\Lambda_{TC}$.
Suppose the case that $\Lambda_{TC} \ll M \ll \Lambda$.
Then a chiral condensate
 $\langle {\bar \psi} \psi \rangle \simeq \Lambda_{TC}^3$
 forms and a global U$(1)_A$ symmetry (anomalous)
 is spontaneously broken,
 where U$(1)_A$ transformation
 is the axial phase transformation of $\psi$.
We have a pseudo-Nambu-Goldstone boson
 $\eta_{TC} \sim {\bar \psi} i \gamma_5 \psi$.

The effect of this technicolor dynamics to the Higgs potential
 can be estimated in several ways.
The most intuitive understanding is obtained
 by considering technifermion one-loop contribution
 in four-dimensional effective theory.
The contribution to the Higgs mass squared is given by
\begin{eqnarray}
 m_{\Phi}^2 &=&
   N_{TC} {\rm tr} \int {{d^4k} \over {(2\pi)^4 i}}
   g_2
   {{\Sigma(-k^2) + k^\mu \gamma_\mu} \over {\Sigma(-k^2)^2 - k^2}}
   g_2
   {{M + k^\nu \gamma_\nu} \over {M^2 - k^2}}
\nonumber\\
 &=&
     4 N_{TC} g_2^2 \int {{d^4k_E} \over {(2\pi)^4}}
    {{\Sigma(k_E^2) M}
     \over
    {(\Sigma(k_E^2)^2 + k_E^2)(M^2 + k_E^2)}}
\nonumber\\
&& 
   - 4 N_{TC} g_2^2 \int {{d^4k_E} \over {(2\pi)^4}}
    {{k_E^2}
     \over
    {(\Sigma(k_E^2)^2 + k_E^2)(M^2 + k_E^2)}},
\label{one-loop-mass}   
\end{eqnarray}
 where $\Sigma$ is the mass function of $\psi$
 which is generated through the technicolor dynamics,
 and the subscription $E$ denotes Euclidean momentum.
The quadratically divergent contribution of the second term
 becomes finite correction in five-dimensional theory
 by virtue of the gauge symmetry.
In the language of the four-dimensional effective theory,
 it is ``regularized'' by infinite Kaluza-Klein modes.
The magnitude of this contribution
 is the same order of the contribution by gauge field,
 but the sign is opposite.
The first term is the contribution of the technicolor dynamics.
The mass function $\Sigma(k_E^2)$
 roughly takes constant value of the order of $\Lambda_{TC}$
 in the region of $k_E^2 < \Lambda_{TC}^2$,
 and quickly decays to zero
 in the region of $k_E^2 > \Lambda_{TC}^2$,
 because of the asymptotically free nature of technicolor.
It is easily understood that
 the sign of the mass function is negative in our case
 due to the effect of the Yukawa interaction
 of Eq.(\ref{Yukawa-interaction})
 by using Cornwall-Jackiw-Tomboulis effective action\cite{CJT}.
In case of $\Lambda_{TC} \ll M$,
 the integral can be estimated as
\begin{equation}
 m_\Phi^2 \vert_{TC}^{\rm one-looop} \simeq
  - {{g_2^2 N_{TC}} \over {8 \pi^2}} {{\Lambda_{TC}^3} \over M}
\end{equation}
 up to some logarithmic correction.
The same result is obtained
 through the mean field approximation
 in the induced higher-dimensional interaction
 by the tree-level exchange of the heavy technifermion $\chi$:
\begin{equation}
 {\cal L}_{eff}
  = {{g_2^2} \over M} {\bar \psi} \psi \Phi^\dag \Phi
  \ \rightarrow \
    {{g_2^2} \over M} \Lambda_{TC}^3 \Phi^\dag \Phi,
\label{induced-interaction}
\end{equation}
 where we define $\Lambda_{TC}$
 by $\Lambda_{TC}^3 \equiv \langle {\bar \psi} \psi \rangle$.
The induced Higgs mass by the technicolor dynamics is now
\begin{equation}
 m_\Phi^2 \vert_{TC} = - g_2^2 {{\Lambda_{TC}^3} \over M}.
\label{Higgs-mass-TC}
\end{equation}

This contribution of the technicolor dynamics
 should be considered with other one-loop contributions
 of gauge bosons and other matter fields in the bulk.
The absolute magnitude of the contribution of the bulk fields
 (including technifermions) are typically larger than
 the technicolor contribution of Eq.(\ref{Higgs-mass-TC}).
In usual gauge-Higgs unification models
 the appropriate value of Higgs mass is expected to be generated
 through the cancellations among the contributions
 of the order of $\Lambda^2 \sim 1 \ \mbox{TeV}^2$
 from many bulk fields.
It is expected that
 nature has chosen an appropriate
 field contents and their orbifold parities
 to realize electroweak symmetry breaking at the weak scale.
Although
 this is certainly a general problem in gauge-Higgs unification model,
 solving this problem and constructing some realistic models
 without this difficulty are not the aim of this paper.
Here,
 we simply imagine a model in which
 these contributions to the Higgs mass are cancelled out
 or smaller than weak scale due to some special fields contents.
In this case the electroweak symmetry breaking scale
 is determined by the dynamics of technifermions.
The actual Higgs vacuum expectation value
 is determined by the Higgs mass squared
 and the Higgs quartic coupling constant $\lambda$
 as $v = \sqrt{-m^2/\lambda} \simeq 250$GeV.
The quartic coupling is also induced
 by one-loop quantum effect of gauge bosons (negative contributions)
 and fermions (positive contributions) in the bulk.
Note that
 the one-loop positive contribution of the technifermion
 is roughly proportional to $N_{TC}$,
 though some non-perturbative correction by
 strong coupling technicolor is expected.
Therefore, large $N_{TC}$ results heavy Higgs.
This may give a solution to
 the general problem of too light Higgs boson
 in usual gauge-Higgs unification models.

In case that
 the technicolor contribution is not dominant in Higgs mass squared,
 the electroweak symmetry breaking scale should be determined
 by the collaboration of the gauge and matter fields
 in the bulk\cite{Kudo-Lim-Yamashita,HHKY}
 as well as the technicolor contribution.
In this case technicolor is not necessary,
 but if it exists there must be a extra field,
 a pseudo-Nambu-Goldstone boson,
 which couples with the Higgs particle in a specific way.
For large $N_{TC}$
 the Higgs mass in case with technicolor can be heavier
 than that in case without technicolor.

\section{Phenomenology}
\label{sec:phenomenology}

Since the technicolor dynamics breaks U$(1)_A$ symmetry,
 which is explicitly broken by the technicolor anomaly,
 there appears massive pseudo-Nambu-Goldstone boson
 $\eta_{TC} \sim {\bar \psi} i \gamma_5 \psi$. 
The scalar resonances which couple with the operator
 ${\bar \psi} \psi$ ($\sigma_{TC}$, for example)
 can couple with two $\eta_{TC}$.
Therefore,
 through the interaction of Eq.(\ref{induced-interaction}),
 we expect the effective interaction
\begin{equation}
 {\cal L}_{eff}^{HH\eta\eta}
  \sim {{g_2^2} \over {M \Lambda_{TC}}}
       \Phi^\dag \Phi \partial^\mu \eta_{TC} \partial_\mu \eta_{TC},
\end{equation}
 where derivative coupling
 is required by chiral U$(1)_A$ symmetry\cite{HSS}.
If $\eta_{TC}$ is lighter than the half of the Higgs mass,
 Higgs can decay into two $\eta_{TC}$
 through the following characteristic interaction
\begin{equation}
 {\cal L}_{eff}^{h\eta\eta}
  \sim {{g_2^2 v} \over {M \Lambda_{TC}}}
       h \partial^\mu \eta_{TC} \partial_\mu \eta_{TC},
\end{equation}
 where $h$ is the physical Higgs field.
If $\eta_{TC}$ is heavy,
 it is associatively produced in pair
 through the Higgs and weak gauge boson production processes.

The precise estimation of the mass of $\eta_{TC}$ is not easy,
 since we have to quantitatively understand the instanton effect
 \cite{'tHooft}.
The very naive scaling-up of QCD gives
\begin{equation}
 {{m_{\eta_{TC}}} \over {\Lambda_{TC}}}
  \simeq {{m_\eta} \over {\Lambda_{QCD}}}
  \simeq 2.5,
\end{equation}
 where we used the result of ``naive dimensional analysis''
 \cite{Manohar-Georgi,Georgi-Randall},
 $\Lambda_{QCD}^3 \equiv
  \langle {\bar q} q \rangle \simeq 4 \pi f_\pi^3$,
 with $f_\pi \simeq 93$MeV and $m_\eta \simeq 550$MeV.
The value of $\Lambda_{TC}$ is model dependent.
In case that
 the technicolor effect dominates the electroweak symmetry breaking,
 $\Lambda_{TC} > 1$TeV and $\eta_{TC}$ is heavy.
Note that larger value of $\Lambda_{TC}$
 means larger value of $M$ (see Eq.(\ref{Higgs-mass-TC})).
In case that
 the technicolor effect is subdominant,
 $\eta_{TC}$ can be lighter.  

Since $\eta_{TC}$ is the lightest ``technihadron''
 and singlet under the standard model gauge group
 (the constituent technifermions are also singlet),
 it is almost stable.
It can decay into two photons only through the three loop effect.
Therefore,
 $\eta_{TC}$ can be a candidate of the cold dark matter
 (further analysis, which is model dependent, is required).

There should be
 a heavy ``technimeson'' $\Theta \sim {\bar \psi} \chi$,
 which can mix with Higgs doublet fields
 through the interaction of Eq.(\ref{Yukawa-interaction}).
The mixing mass matrix can be roughly estimated as
\begin{equation}
 {\cal L}_{\Phi\Theta} =
 -
 \left(
  \begin{array}{cc}
   \Phi^\dag & \Theta^\dag
  \end{array}
 \right)
 \left(
  \begin{array}{cc}
   0 & g_2 \Lambda_{TC} \sqrt{M \Lambda_{TC}} \\
   g_2 \Lambda_{TC} \sqrt{M \Lambda_{TC}} & M^2
  \end{array}
 \right)
 \left(
  \begin{array}{c}
   \Phi \\ \Theta
  \end{array}
 \right)
\end{equation}
 using the technique of the heavy quark effective theory.
Precisely saying, we have two Higgs doublet fields.
In case $\Lambda_{TC} \ll M$,
 the mixing angle is small and
 the heavy Higgs doublet field decouples.
It is interesting to note that
 we have negative mass squared of the Higgs doublet field
 of the order of $-g_2^2 \Lambda_{TC}^3/M$ from this mass matrix
 of ``bosonic see-saw'' form\cite{Calmet,Kim}.
Since $\Theta$ is much heavier than $\Lambda_{TC}$,
 this result in the low-energy effective theory
 may be thought as the same result of Eq.(\ref{Higgs-mass-TC})
 in the high-energy fundamental theory.

\section{Conclusions}
\label{sec:conclusions}

We have proposed
 a scenario of electroweak symmetry breaking
 which is applicable in case that
 there is no tree-level potential of Higgs doublet field.
The gauge-Higgs unification model is an example,
 in which the tree-level potential of Higgs doublet field
 is forbidden by gauge symmetry.
The strong coupling ``technicolor'' dynamics
 can play a role to set electroweak symmetry breaking scale
 through the Yukawa couplings among
 heavy and light ``technifermions''
 and elementary Higgs doublet field.
The prediction of this scenario
 is a pseudo-Nambu-Goldstone boson,
 which is singlet under the standard model gauge symmetry,
 but interacts with Higgs doublet field in a specific way.
We leave concrete model buildings and phenomenological analysis
 for future works.

\section*{Acknowledgment}

N.~H. and N.~K. thank the theory division of KEK for hospitality.
N.~H. is supported in part
 by the Grand-in-Aid for Scientific Research
 \#16028214 and \#16540258.
N.~K. and N.~O. are supported in part
 by the Grand-in-Aid for Scientific Research
 \#16340078 and \#15740164, respectively.


\begin{thebibliography}{}
\bibitem{Weinberg}
 S.~Weinberg, Phys. Rev., D13 (1976) 974.
\bibitem{Susskind}
 L.~Susskind, Phys. Rev., D20 (1979) 2619.
\bibitem{Peskin-Takeuchi}
 M.~E.~Peskin and T.~Takeuchi,
  Phys. Rev. Lett. 65 (1990) 964; Phys. Rev. D46 (1992) 381.
\bibitem{Sundram-Hsu}
 R.~Sundrum and S.~D.~H.~Hsu, Nucl. Phys. B391 (1993) 127.
\bibitem{Kitazawa-Yanagida}
 N.~Kitazawa and T.~Yanagida, Phys. Lett. B383 (1996) 78.
\bibitem{Appelquist-Sannino}
 T.~Appelquist and F.~Sannino, Phys. Rev. D59 (1999) 067702.
\bibitem{HHS}
 D.~K.~Hong, S.~D.~H.~Hsu and F.~Sannino,
 Phys. Lett. B597 (2004) 89.
%
\bibitem{Manton}
 N.~S.~Manton, Nucl. Phys. B158 (1979) 141.
\bibitem{Fairlie}
 D.~B.~Fairlie, J. Phys. G5 (1979) L55; Phys. Lett. B82 (1979) 97.
\bibitem{Hosotani}
 Y.~Hosotani, Phys. Lett. B126 (1983) 309;
 Phys. Lett. B129 (1984) 193;
 Phys. Rev. D29 (1984) 731;
 Ann. of Phys. 190 (1989) 233.
\bibitem{Hall-Nomura-Smith}
 L.~J.~Hall, Y.~Nomura and D.~R.~Smith,
 Nucl. Phys. B639 (2002) 307.
\bibitem{Burdman-Nomura}
 G.~Burdman and Y.~Nomura,
 Nucl. Phys. B656 (2003) 3.
\bibitem{Kudo-Lim-Yamashita}
 M.~Kubo, C.~S.~Lim and H.~Yamashita,
 Mod. Phys. Lett. A17 (2002) 2249.
\bibitem{HHKY}
 N.~Haba, Y.~Hosotani, Y.~Kawamura and T.~Yamashita,
 Phys. Rev. D70 (2004) 015010.
%
\bibitem{Kaplan-Georgi}
 D.~B.~Kaplan and H.~Georgi, Phys. Lett. B136 (1984) 183.
\bibitem{Choi-Kim}
 K.~Choi and J.~E.~Kim, Phys. Lett. B159 (1985) 131.
\bibitem{little-Higgs}
 M.~Schmaltz and D.~Tucker-Smith,
 hep-ph/0502182, and references there in.
\bibitem{CJT}
 J.~M.~Cornwall, R.~Jackiw and E.~Tomboulis,
 Phys. Rev. D10 (1974) 2428.
\bibitem{HSS}
 M.~Harada, F.~Sannino and J.~Schechter,
 Phys. Rev. D54 (1996) 1991.
\bibitem{'tHooft}
 G.~'t Hooft, Phys. Rept. 142 (1986) 357.
\bibitem{Manohar-Georgi}
 A.~Manohar and H.~Georgi, Nucl. Phys. B234 (1984) 189.
\bibitem{Georgi-Randall}
 H.~Georgi and L.~Randall, Nucl. Phys. B276 (1986) 241.
\bibitem{Calmet}
 X.~Calmet, Eur. Phys. J. C28 (2003) 451.
\bibitem{Kim}
 H.~D.~Kim, hep-ph/0501059.
\end{thebibliography}
\end{document}